\documentclass[twocolumn,amsmath,amssymb]{revtex4}
\usepackage{graphicx}
\usepackage{dcolumn}
\usepackage{bm}

\newcommand{\bea}{\begin{eqnarray}}
\newcommand{\eea}{\end{eqnarray}} \newcommand{\bg}{\begin{gather}}
\newcommand{\eg}{\end{gather}}
\newcommand{\bseq}{\begin{subequations}}
\newcommand{\eseq}{\end{subequations}}

\newcommand{\be}{\begin{eqnarray}}
\newcommand{\beq}{\begin{equation}}
\newcommand{\eeq}{\end{equation}}
\newcommand{\ee}{\end{eqnarray}}

\newcommand{\bmp}{\noindent\begin{minipage}{16cm}}
\newcommand{\emp}{\end{minipage}\vskip 7mm} 

\usepackage{graphicx}
\usepackage{dcolumn}
\usepackage{bm}
\usepackage{amsmath}
\usepackage{amsfonts}
\begin{document}

\title{Limits on Self-Interacting Dark Matter}
\author{Chris {\sc Kouvaris}}\email{kouvaris@cp3.sdu.dk}
\affiliation{$\text{CP}^3$-Origins \& Danish Institute for Advanced Study DIAS, \\ University of Southern Denmark, Campusvej 55, Odense 5230, Denmark}
\begin{abstract}
We impose new severe constraints on the self-interactions of fermionic asymmetric dark matter based on observations of nearby old neutron stars. WIMP self-interactions mediated by Yukawa-type interactions can lower significantly the number of WIMPs necessary for gravitational collapse of the WIMP population accumulated in a neutron star. Even nearby neutron stars located at regions of low dark matter density can accrete sufficient number of WIMPs 
that can potentially collapse, form a mini black hole, and destroy the host star. Based on this, we derive constraints on the WIMP self-interactions which in some cases are by several orders of magnitude stricter than the ones from the bullet cluster (which are currently considered the most stringent).
 \\[.1cm]
{\footnotesize  \it Preprint: CP$^3$-Origins-2011-041 \& DIAS-2011-36.}\end{abstract}

\maketitle 

It is highly possible that dark matter (DM) is in the form of Weakly Interacting Massive Particles (WIMPs). Although the cross section in WIMP-nucleon interactions is severely constrained in underground detectors, we have little information regarding self-interactions among WIMPs other than gravitational attraction.  The possibility of WIMP self-interactions has been studied in the past~\cite{deLaix:1995vi}, and in principle there is no argument against it. On the contrary WIMP self-interactions have been proposed as a solution of  the discrepancy between simulations of collisionless DM and observations of galaxies~\cite{Spergel:1999mh,Loeb:2010gj}.
Simulations of collisionless cold DM predict a cusped profile with a steep power law-like behavior in the DM density ~\cite{Navarro:2008kc}. This picture is contradicted by observations of the rotation curves of gas-rich dwarf galaxies. Such galaxies are dominated by DM and therefore it is expected to give more reliable information regarding the DM density profile. Observations of such dwarf galaxies suggest that their DM halos have an approximately constant density rather than a power law scaling. Generally speaking self-interactions can alleviate this core/cusp problem. Collisions among WIMPs can potentially flatten out the cusp, thus creating the almost constant density favored by observations. Such WIMP self-interactions should in general be strong enough in order to have frequent WIMP-WIMP collisions at the core, but they have to be short range interactions  in order to avoid changing large scale structure~\cite{Gradwohl:1992ue}. In addition, for such a scenario to be realized, WIMP self-annihilations should be either absent or restricted.

Although asymmetric dark mater is not a new idea~\cite{Nussinov:1985xr,Barr:1990ca,Gudnason:2006ug,Gudnason:2006yj}, it has recently attracted a lot of  interest~\cite{ Foadi:2008qv,Khlopov:2007ic,Khlopov:2008ty,Dietrich:2006cm,Sannino:2009za,Ryttov:2008xe,Sannino:2008nv,Kaplan:2009ag,Frandsen:2009mi,MarchRussell:2011fi,Frandsen:2011cg,Gao:2011ka,Arina:2011cu,Buckley:2011ye,Lewis:2011zb,Davoudiasl:2011fj,Graesser:2011wi,Bell:2011tn}. This is because it can be an alternative to thermally produced WIMP DM, it can link the asymmetry in the dark sector to the baryonic asymmetry, and it can possibly reconcile discrepancies among underground direct search experiments through interference of different channels~\cite{DelNobile:2011je,DelNobile:2011yb}. In an asymmetric DM scenario, annihilations are rare due to the fact that most of anti-WIMPs have already been annihilated in the past. Therefore self-interacting asymmetric DM is quite an attractive possibility for resolving the cusp/core problem. 

In this letter we impose model-independent constraints on self-interacting asymmetric DM candidates with Yukawa-type interactions based on neutron star observations. Compact stars such as neutron stars and white dwarfs can potentially constrain WIMP masses and cross sections with nuclei~\cite{Goldman:1989nd,Kouvaris:2007ay,Bertone:2007ae,Sandin:2008db,McCullough:2010ai,Kouvaris:2010vv,Casanellas:2010sj,deLavallaz:2010wp,Kouvaris:2010jy,Ciarcelluti:2010ji, Kouvaris:2011fi,McDermott:2011jp}. For asymmetric DM, such constraints are based on the fact that if enough WIMPs have been accreted and trapped within the compact star, they can collapse forming a black hole that eventually destroy the host star. Therefore the existence of relatively old compact stars in our Galaxy in regions of sufficiently rich DM density can exclude a lot of phase space of asymmetric DM candidates. Based on such arguments, spin-dependent interactions of asymmetric WIMPs~\cite{Kouvaris:2010jy}, and spin-(in)dependent interactions of bosonic asymmetric DM~\cite{Kouvaris:2011fi,McDermott:2011jp} have been severely constrained. However, in the case of fermionic DM, constraints based on the above argument are very weak. In order for WIMPs to collapse and overcome the WIMP Fermi pressure, a large number of trapped WIMPs is required, making the constraints valid only for very large (and uninteresting) WIMP masses. However, the existence of attractive self-interactions tends to alleviate the effect of Fermi pressure, changing drastically the number of WIMPs  required for collapse. In this letter we assume that DM WIMPs (of mass $m$) self-interact via a Yukawa-type interaction of the form $\alpha \phi \bar{\psi}\psi$. If $\phi$ (with a mass $\mu$) is a scalar, such self-interactions are always attractive, inducing a short range ``dark force'' with a potential $V(r) =-\alpha \exp [-\mu r]/r$. This is the type of WIMP-self-interactions that have been proposed as an explanation of the flatness of DM density in dwarf galaxies~\cite{Loeb:2010gj}, or in order to fit the direct DM search experimental results~\cite{Fornengo:2011sz}.

WIMPs passing through a neutron star can scatter and potentially be trapped inside. The accretion of WIMPs onto a typical 1.4$M_{\bigodot}$ 10 km neutron star   taking into account relativistic effects has been calculated in~\cite{Kouvaris:2007ay,Kouvaris:2010vv}. The number of accumulated WIMPs is
\begin{equation}	
N_{acc}=4.3 \times 10^{46} 
\left (\frac{\rho_{\text{dm}}}{10^3 \text{GeV}/\text{cm}^3} \right ) \left (\frac{\text{GeV}}{m} \right  ) \left (\frac{20\text{km/s}}{v} \right ) t_9 f, \label{rate}  \end{equation} where $v$ is the WIMP average velocity, $t_9$ is the time in billion years, and $f$ is  a factor which is 1 if the WIMP-nucleon cross section $\sigma > 10^{-45}\text{cm}^2$, and $\sigma/(10^{-45}\text{cm}^2)$ if  $\sigma < 10^{-45}\text{cm}^2$. Once the WIMPs get trapped, they thermalize  with nucleons acquiring the same temperature in a very short time interval~\cite{Kouvaris:2010vv} and follow a Fermi-Dirac distribution.  The WIMP thermal radius can be  determined by applying the virial theorem for a random WIMP $i$
\beq 2\langle E_k \rangle=\frac{8}{3}\pi G\rho m r^2+\frac{GNm^2}{r}+ \left \langle \sum_j \alpha\frac{e^{-\mu r_{ij}}}{r_{ij}}+\alpha \mu e^{-\mu r_{ij}} \right \rangle, \label{virial} \eeq where $\langle E_k \rangle$ is the average kinetic energy of a single WIMP, $N$ is the total number of WIMPs, $\rho$ is the core mass density of the neutron star, $r$ is the WIMP thermal radius, and $r_{ij}$ is the distance between WIMP $i$ and WIMP $j$. The sum runs over all WIMPs (except $i$). All quantities have been averaged over time and $i$. The virial theorem for a generic potential  of the form $V(r)=r^ne^{-ar^p}$ is $2\langle E_k \rangle=n\langle V \rangle -ap \langle r^p V \rangle $~\cite{virial}. It can be easily seen that the first two terms on the right hand side correspond to the gravitational potential energy due to the core of the neutron star and the self-gravitation of WIMPs respectively. The last two terms correspond to the average attractive Yukawa potential energy. Since most of the WIMPs are concentrated within the thermal radius $r$, if $x$ is the mean distance between two closest neighbor WIMPs, $x=r/N^{1/3}$. At the beginning, as WIMPs start accumulating in the star, the first term of Eq.~(\ref{virial}) is the dominant one. As $N$ increases, the other terms start to become more important. It is clear that Yukawa forces can play a significant role if $\mu x<<1$ or 
\beq \mu r/N^{1/3}<<1. \label{satur} \eeq In this case, the exponential will be roughly close to $\sim1$ and the Yukawa force between two WIMPs will scale as a Coulomb one.  In order to determine the thermal radius, one has to 
solve Eq.~(\ref{virial}) for $r$ keeping in mind that the kinetic energy $E_k=(3/2)k_BT$ for nondegenerate WIMPs, or $E_k=(9 \pi/4)^{2/3}N^{2/3}/(2mr^2)$ (for nonrelativistic degenerate WIMPs) once $n\lambda^3>2$, where $n$ is the WIMP number density, and $\lambda =h/\sqrt{2 \pi mk_B T}$. We assumed for simplicity that WIMPs have two degenerate states (spin-up spin-down). The generalization to a higher number of states is straightforward. For typical parameters of a neutron star (and a thermal radius $r_{th}=[9k_BT/(8 \pi G \rho m)]^{1/2}= 220\text{cm}/\sqrt{m}$~\cite{Kouvaris:2011fi}), WIMPs become degenerate once their number is $N_{deg}=6\times10^{35}$. There are two different possibilities as the number of trapped WIMPs increases depending on the parameters of the Yukawa forces. Yukawa self-interactions might dominate Eq.~(\ref{virial}) either before or after the degenerate limit has been reached. 

{\it Self-attraction after degeneracy.}  The kinetic energy is the one of a nonrelativistic degenerate gas, and therefore the thermal radius is
\beq r_{\text{th}}= \left (\frac{9 \pi N}{4} \right )^{1/6} \left (\frac{3}{8\pi G \rho m^2} \right )^{1/4} \simeq 1.14 \times 10^{10} \frac{N^{1/6}}{m^{1/2}} \text{GeV}^{-1}. \eeq 
We have used a typical value $\rho= 5 \times 10^{38} \text{GeV}/\text{cm}^3$. There are two possibilities here: Either self-attraction (3rd and 4th terms in Eq.~(\ref{virial})) dominates after Eq.~(\ref{satur}) is satisfied, or not.  In the first case, a comparison between the 1st and the sum of the 3rd and 4th term of Eq.~(\ref{virial}) shows that self-attraction takes place after the saturation of Eq.~(\ref{satur}) condition, if $\alpha<4.38 \mu/m$. At this point we define the quantity $y=\mu r/N^{1/3}$ which will be more convenient to work with instead of dealing with $r$. The particles become relativistic once the momentum becomes comparable to the mass $(9 \pi/4)^{1/3} N^{1/3}/r \simeq m$, or $y_{\text{rel}}=(9 \pi/4)^{1/3} \mu /m.$  Now we have to estimate the potential energy due to  the Yukawa interactions for $y<1$. The energy is \beq E_i= \alpha \sum_j \frac{e^{-\mu r_{ij}}}{r_{ij}}, \eeq where $i$ denotes a specific particle in question and $j$ runs over all the rest. Apparently if $y>1$, due to the exponential decay of the potential with separation distance between particles, the most significant contribution comes from the nearest neighbor. However if $y<1$, not only there is no exponential suppression for the  nearest neighbor, but also more distant particles can contribute as long as $r_{ij}<1/\mu$. It is easy to show that for $y<1$, the number of particles inside a sphere with radius $1/\mu$ is $1/y^3$. This is the number of particles that will contribute significantly to the potential energy. If we take an approximate average $\langle1/r_{ij} \rangle \simeq \mu$, the Yukawa potential energy is $E_i=\alpha \mu/y^3$.

\begin{figure}[h!]
\begin{center}
\includegraphics[width=.39
\textwidth, height=0.3 \textwidth
]{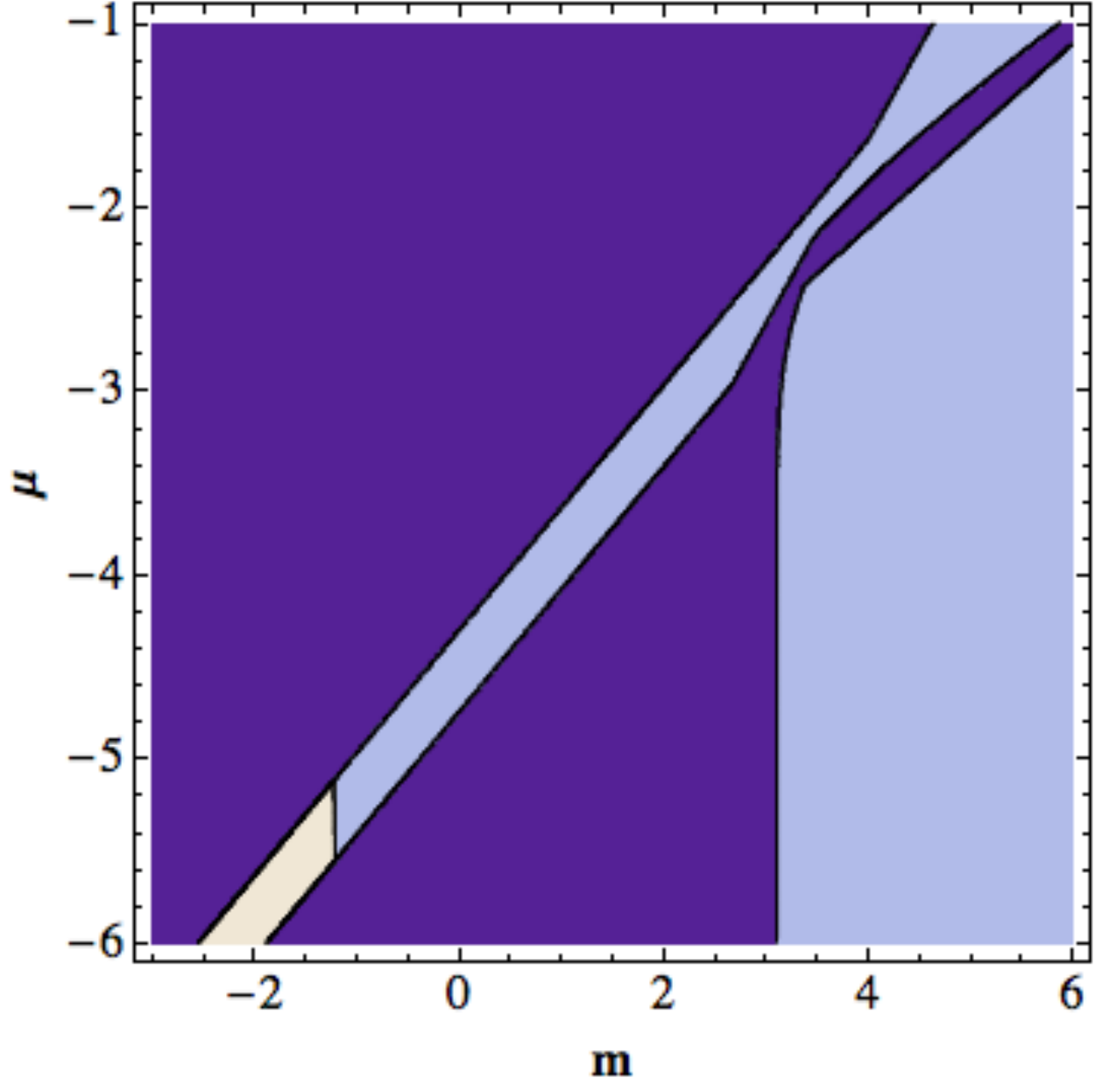}
\includegraphics[width=.4\textwidth, height=0.3 \textwidth
]{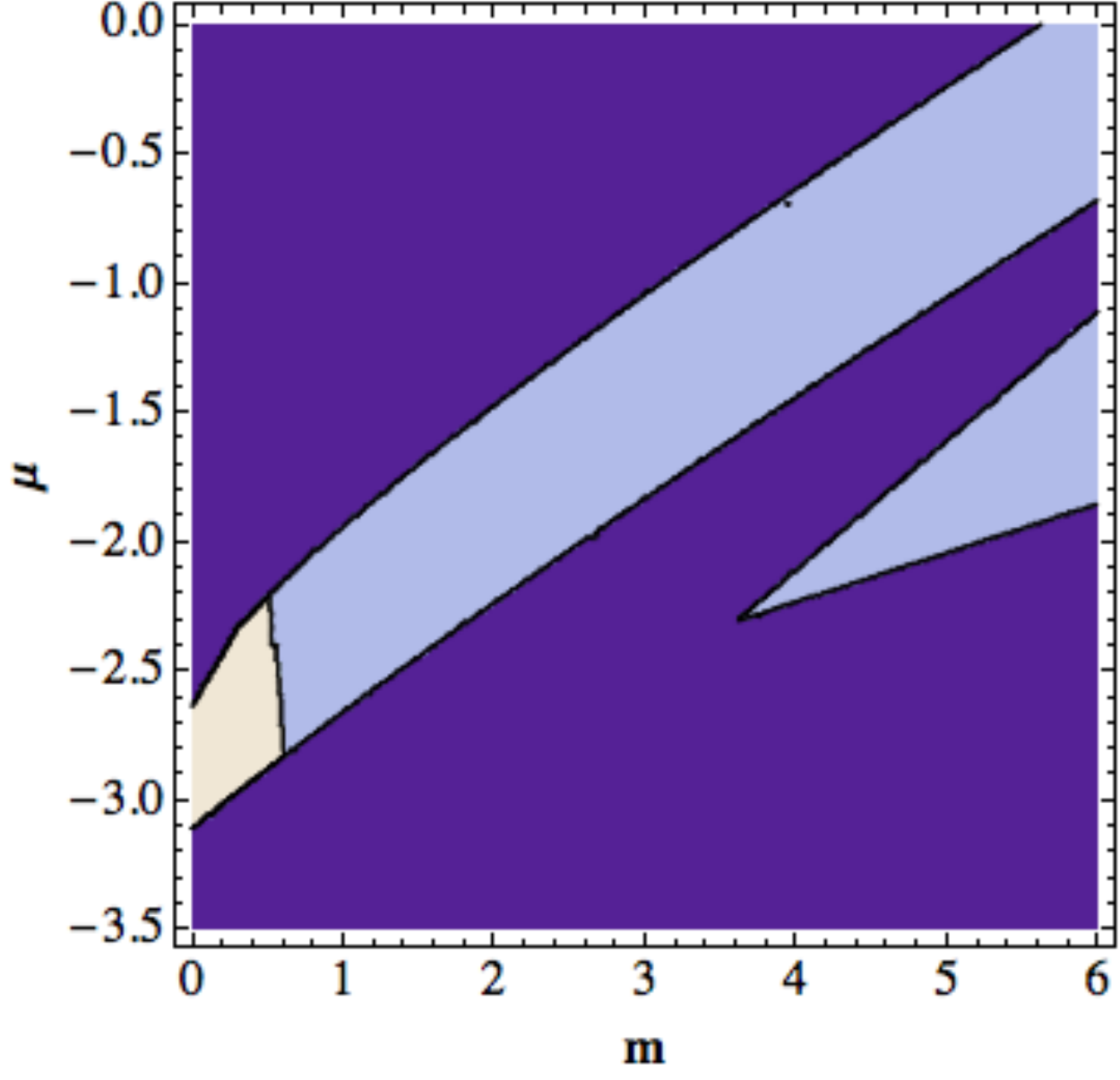}
\includegraphics[width=.4\textwidth, height=0.3 \textwidth
]{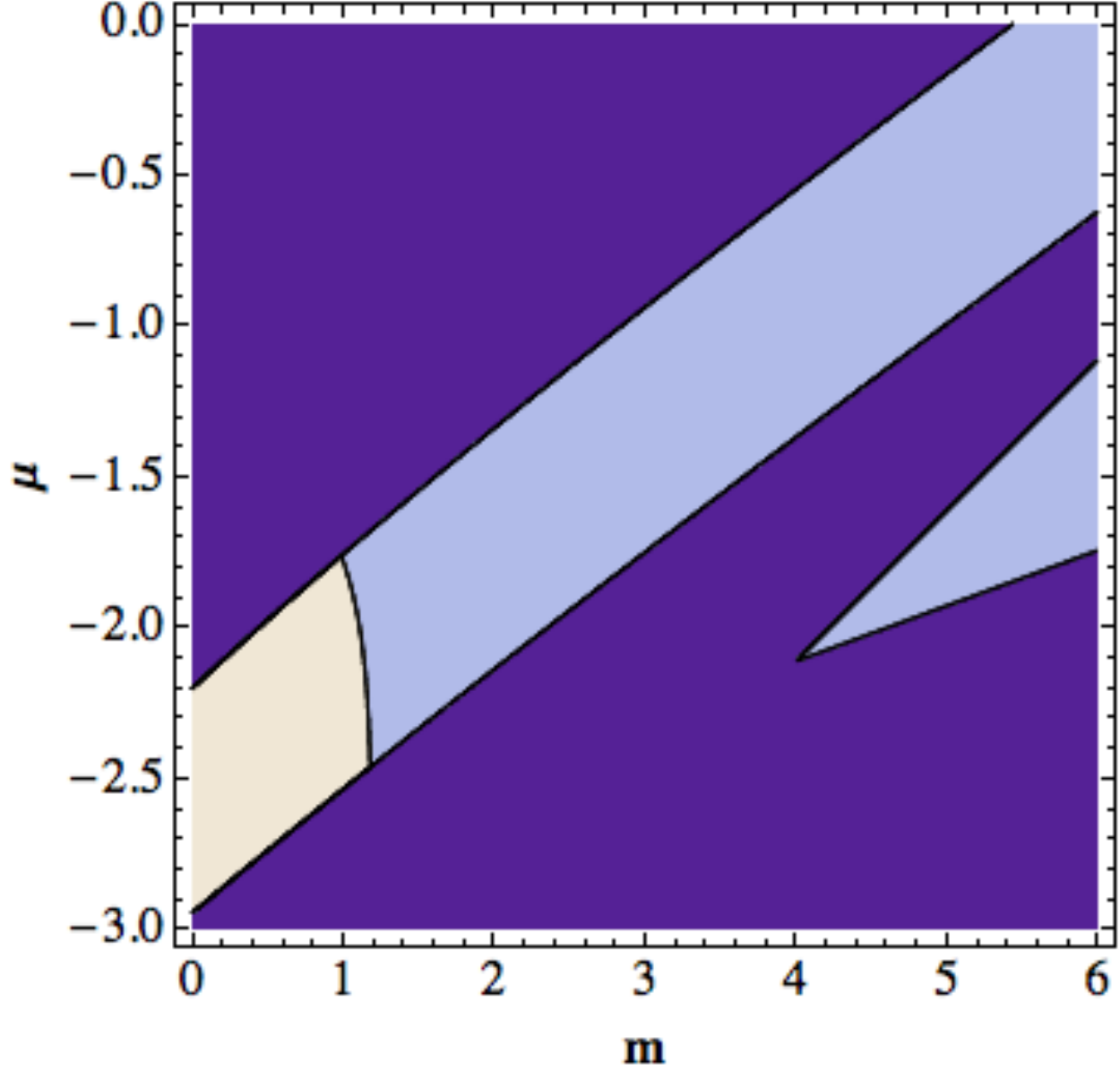}\caption{Excluded regions (in light blue) for the WIMP mass $m$ and the mediator mass $\mu$ in logarithmic scales of GeV for three different Yukawa couplings $\alpha=10^{-5}$, 0.01, and 0.1 from top to bottom. Light brown regions are also excluded by constraints on WIMP self interactions based on the bullet cluster. The larger exclusion region in plot 1 and the ``triangle'' regions in plots 2 and 3 correspond to gravitational collapse initiated before WIMPs have become degenerate.}
\end{center}
\end{figure}

  As we mentioned above, for $\alpha<4.38 \mu/m$, self-attraction dominates after Eq.~(\ref{satur}) has been saturated.  There are two possible cases here. The first one is to have self-attraction dominating before the particles become relativistic. In this case the Yukawa self-attraction overcomes nonrelativistic Fermi pressure once \beq \frac{\alpha\mu}{y^3}>\left ( \frac{9\pi}{4} \right )^{2/3}\frac{\mu^2}{m y^2}.\eeq  This happens when $y<y_1=\alpha m/3.65\mu$.  As the WIMPs keep falling, they become eventually relativistic. In order to further collapse, self-attraction should win over the relativistic Fermi pressure. This means \beq  \frac{\alpha\mu}{y_{rel}^3}>\left ( \frac{9\pi}{4} \right )^{1/3}\frac{\mu}{ y_{rel}} \Leftrightarrow \alpha >7 \frac{\mu^2}{m^2}.\eeq   Therefore for $7 \mu^2/m^2< \alpha <4.38 \mu/m$, it is guaranteed that WIMPs can gravitationally collapse forming a black hole so long $y<y_1$. This condition on $y$ can be transformed to a condition on the  number of accreted WIMPs through Eq.~(4). 
  In this letter we are going to consider limits derived only from nearby old neutron stars with DM densities of $\rho_{dm}=0.3 \text{GeV}/\text{cm}^3$.  Globular clusters can give stricter constraints, but we have chosen here to present results that do not depend on the not well known DM density in globular clusters, but rather on the unambiguous low densities of nearby neutron stars. There is one extra constraint that we have to consider.  If the above conditions are satisfied, WIMPs will form a black hole. However, as it was estimated in~\cite{Kouvaris:2011fi}, only black holes with masses larger than $5.7\times 10^{36}$ GeV can eventually destroy the host star. Smaller black holes decay due to the fact that Hawking evaporation  dominates over the black hole accretion of matter in the neutron star.
  
The case where Yukawa forces dominate at a $y<y_{rel}$ can also in principle lead to formation of a black hole. However, we found that this happens for $\alpha<7\mu^2/m^2$. Practically for such small couplings, there is not enough phase space, meaning that accretion of WIMPs in nearby neutron stars in billions years is not enough to meet the condition for collapse.

 We move now to the case where $\alpha> 4.38 \mu/m$. This means that Yukawa self-attraction becomes important before $y=1$. This is distinctively different from the previous case because  the Yukawa potential here is still exponentially suppressed (and has $1/y$ instead of $1/y^3$ dependence). The virial theorem Eq.~(\ref{virial}) dictates the thermal radius of the WIMP sphere. Ignoring WIMP self-gravitation which is small for our case of interest, Eq.~(\ref{virial}) can be rewritten as
\beq \left ( \frac{9 \pi}{4} \right )^{2/3} \frac{\mu}{m}\frac{1}{y^2}=\alpha \left (1 + \frac{1}{y} \right )e^{-y}+\frac{8}{3}\pi G\rho \frac{m}{\mu^3}N^{2/3}y^2. \label{virial2} \eeq  Note that in this case we deal with $y>1$ and therefore we have approximated the sum over all WIMPs in Eq.~(\ref{virial}) simply by the closest neighbor. There are up to three solutions for this equation in the case where $\alpha >4.38 \mu/m$. One of them lies at $y<1$ and the rest at $y>1$. As $N$ increases, the two solutions with $y>1$ come closer to each other. Eventually at a critical value $N_{crit}$, the two solutions coincide and then disappear leaving only the solution with $y<1$. We have solved numerically Eq.~(\ref{virial2}) to identify $N_{crit}$ as well as the value of $y$ that is a solution when $N$ is slighly larger than $N_{crit}$. We have compared the Yukawa potential energy to the Fermi one, and as before the constraint that we derive is that $\alpha>7 \mu^2/m^2$. We should also mention here that the same result can be obtained by minimizing the energy of the system. One can easily check that the minimization of the energy of the system gives the virial Eq.~(\ref{virial2}).

In all of the above constraints, we have implicitly assume that $y_{rel}<1$ (or equivalently  $\mu<0.52 m$) i.e. WIMPs become relativistic when already Yukawa forces are not exponentially suppressed. In the opposite case where  $y_{rel}>1$, WIMPs become relativistic before Yukawa forces become effective. In this case we find that there is a small phase space to be excluded, but it is rather uninteresting since it lies on a region excluded already by constraints on WIMP self-interactions based on the bullet cluster. This case can also give constraints on strongly coupled interactions ($\alpha>1$) but in this letter we focus on perturbative couplings $\alpha<1$.
 
{\it Self-attraction before degeneracy.}  Depending on the parameters, it is possible for Yukawa self-attractions to become important before the WIMPs have reached the degeneracy limit which corresponds to $y_{deg}=1. 3 \times 10^4 \mu/ \sqrt{m}$ (both $m$ and $\mu$ measured in GeV). This happens when the last two terms of Eq.~(\ref{virial}) become comparable to the first one. We have identified numerically the parameter region that this happens and we have estimated the number of WIMPs $N_{s}$ needed for that. Note that this number has to be smaller than $N_{deg}$.  In such a case, it is possible to form a WIMP black hole if $y_{deg}<1$ and the Yukawa potential dominates over nonrelativistic and relativistic Fermi pressure. This leads to the following constraints respectively \beq \alpha> (9\pi/4)^{2/3}\frac{\mu}{m} y_{deg}\simeq 4.8 \times 10^4\frac{\mu^2}{m^{3/2}} \quad \&  \quad \alpha> 7\frac{\mu^2}{m^2}, \eeq where for all cases of interest the first one is much tighter and therefore sufficient for collapse. Once more in order to impose actual constraints the WIMP black hole must be larger than  $5.7\times 10^{36}$ GeV in order to grow and destroy the host star. 

 In Fig.~1 we present all the exclusion regions  based on the existence of old neutron stars  in the vicinity of the Earth where the DM density is $\rho= 0.3$ GeV$/\text{cm}^3$, such as J0437-4715~\cite{Kargaltsev:2003eb} and J0108-1431~\cite{Mignani:2008jr} (140 pc and 130 pc from the Earth, respectively). The fact that these stars have not been destroyed by a black hole after accreting WIMPs for at least one billion years, can impose severe constraints on the possible WIMP mass and the mass of the mediator that is exchanged. The constraints are valid as long as the WIMP-nucleon cross section is not smaller than the $\sigma_{crit}=10^{-45}\text{cm}^2$. However, we found that the exclusion regions do not change significantly  with $\sigma$ as low as $ 10^{-48}\text{cm}^2$. Old neutron stars in globular clusters can potentially give stronger constraints, albeit the arbitrariness of the local DM density. In addition, neutron stars in binary systems might increase by a factor of few the dark matter accretion~\cite{Brayeur:2011yw}. In Fig.~1 we present new constraints not excluded by the bullet cluster observations. Using the WIMP-WIMP cross section for Yukawa interactions~\cite{Loeb:2010gj} with a relative velocity of 4500 km/s relevant for the bullet cluster, we exclude a whole new region for self-interactions (for asymmetric WIMPs).  Comparatively, if one were to estimate how stricter are the constraints derived here, a WIMP of mass 1 TeV, exchanging a scalar of 100 MeV with $\alpha=0.01$, would lead to a WIMP-WIMP cross section exclusion of 10 orders of magnitude lower than the  bullet cluster limit $\sigma_{self}/m <2 \times 10^{-24}\text{cm}^2/\text{GeV}$~\cite{Randall:2007ph}.
 
I would like to thank P. Tinyakov for useful comments on the manuscript.


\begin{thebibliography}{99}	

\bibitem{deLaix:1995vi}
  A.~A.~de Laix, R.~J.~Scherrer, R.~K.~Schaefer,
  Astrophys.\ J.\  {\bf 452}, 495 (1995).
  [astro-ph/9502087].
  
\bibitem{Spergel:1999mh}
 D.~N.~Spergel and P.~J.~Steinhardt,
 Phys.\ Rev.\ Lett.\  {\bf 84}, 3760 (2000)
 [arXiv:astro-ph/9909386].

\bibitem{Loeb:2010gj}
 A.~Loeb and N.~Weiner,
 Phys.\ Rev.\ Lett.\  {\bf 106}, 171302 (2011)
 [arXiv:1011.6374 [astro-ph.CO]].

\bibitem{Navarro:2008kc}
 J.~F.~Navarro {\it et al.},
 arXiv:0810.1522 [astro-ph].

\bibitem{Gradwohl:1992ue}
 B.~A.~Gradwohl and J.~A.~Frieman,
 Astrophys.\ J.\  {\bf 398}, 407 (1992).

\bibitem{Nussinov:1985xr}
 S.~Nussinov,
 Phys.\ Lett.\  B {\bf 165}, 55 (1985).

\bibitem{Barr:1990ca}
 S.~M.~Barr, R.~S.~Chivukula and E.~Farhi,
 Phys.\ Lett.\  B {\bf 241}, 387 (1990).

\bibitem{Gudnason:2006ug}
 S.~B.~Gudnason, C.~Kouvaris and F.~Sannino,
 Phys.\ Rev.\  D {\bf 73}, 115003 (2006)
 [arXiv:hep-ph/0603014].

\bibitem{Gudnason:2006yj}
 S.~B.~Gudnason, C.~Kouvaris and F.~Sannino,
 Phys.\ Rev.\  D {\bf 74}, 095008 (2006)
 [arXiv:hep-ph/0608055].

\bibitem{Khlopov:2007ic}
  M.~Y.~.Khlopov, C.~Kouvaris,
  Phys.\ Rev.\  {\bf D77}, 065002 (2008).
  [arXiv:0710.2189 [astro-ph]].
  
\bibitem{Foadi:2008qv}
 R.~Foadi, M.~T.~Frandsen and F.~Sannino,
 Phys.\ Rev.\  D {\bf 80}, 037702 (2009)
 [arXiv:0812.3406 [hep-ph]].

\bibitem{Khlopov:2008ty}
 M.~Y.~Khlopov and C.~Kouvaris,
 Phys.\ Rev.\  D {\bf 78}, 065040 (2008)
 [arXiv:0806.1191 [astro-ph]].

\bibitem{Dietrich:2006cm}
 D.~D.~Dietrich and F.~Sannino,
 Phys.\ Rev.\  D {\bf 75}, 085018 (2007)
 [arXiv:hep-ph/0611341].

\bibitem{Sannino:2009za}
 F.~Sannino,
 Acta Phys.\ Polon.\  B {\bf 40}, 3533 (2009)
 [arXiv:0911.0931 [hep-ph]].

\bibitem{Ryttov:2008xe}
 T.~A.~Ryttov and F.~Sannino,
 Phys.\ Rev.\  D {\bf 78}, 115010 (2008)
 [arXiv:0809.0713 [hep-ph]].

\bibitem{Sannino:2008nv}
 F.~Sannino and R.~Zwicky,
 Phys.\ Rev.\  D {\bf 79}, 015016 (2009)
 [arXiv:0810.2686 [hep-ph]].

\bibitem{Kaplan:2009ag}
 D.~E.~Kaplan, M.~A.~Luty and K.~M.~Zurek,
 Phys.\ Rev.\  D {\bf 79}, 115016 (2009)
 [arXiv:0901.4117 [hep-ph]].

\bibitem{Frandsen:2009mi}
 M.~T.~Frandsen and F.~Sannino,
 Phys.\ Rev.\  D {\bf 81}, 097704 (2010)
 [arXiv:0911.1570 [hep-ph]].

\bibitem{MarchRussell:2011fi}
 J.~March-Russell and M.~McCullough,
 arXiv:1106.4319 [hep-ph].

\bibitem{Frandsen:2011cg}
 M.~T.~Frandsen, F.~Kahlhoefer, S.~Sarkar and K.~Schmidt-Hoberg,
 JHEP {\bf 1109}, 128 (2011)
 [arXiv:1107.2118 [hep-ph]].

\bibitem{Gao:2011ka}
 X.~Gao, Z.~Kang and T.~Li,
 arXiv:1107.3529 [hep-ph].

\bibitem{Arina:2011cu}
 C.~Arina and N.~Sahu,
 Nucl.\ Phys.\  B {\bf 854}, 666 (2012)
 [arXiv:1108.3967 [hep-ph]].

\bibitem{Buckley:2011ye}
 M.~R.~Buckley and S.~Profumo,
 arXiv:1109.2164 [hep-ph].

\bibitem{Lewis:2011zb}
 R.~Lewis, C.~Pica and F.~Sannino,
 arXiv:1109.3513 [hep-ph].

\bibitem{Davoudiasl:2011fj}
 H.~Davoudiasl, D.~E.~Morrissey, K.~Sigurdson and S.~Tulin,
 arXiv:1106.4320 [hep-ph].

\bibitem{Graesser:2011wi}
 M.~L.~Graesser, I.~M.~Shoemaker and L.~Vecchi,
 JHEP {\bf 1110}, 110 (2011)
 [arXiv:1103.2771 [hep-ph]].

\bibitem{Bell:2011tn}
 N.~F.~Bell, K.~Petraki, I.~M.~Shoemaker and R.~R.~Volkas,
 arXiv:1105.3730 [hep-ph].

\bibitem{DelNobile:2011je}
 E.~Del Nobile, C.~Kouvaris and F.~Sannino,
 Phys.\ Rev.\  D {\bf 84}, 027301 (2011)
 [arXiv:1105.5431 [hep-ph]].

\bibitem{DelNobile:2011yb}
  E.~Del Nobile, C.~Kouvaris, F.~Sannino, J.~Virkajarvi,
  
  [arXiv:1111.1902 [hep-ph]].
\bibitem{Goldman:1989nd}
 I.~Goldman and S.~Nussinov,
 Phys.\ Rev.\  D {\bf 40}, 3221 (1989).

\bibitem{Kouvaris:2007ay}
 C.~Kouvaris,
 Phys.\ Rev.\  D {\bf 77}, 023006 (2008)
 [arXiv:0708.2362 [astro-ph]].

\bibitem{Bertone:2007ae}
 G.~Bertone and M.~Fairbairn,
 Phys.\ Rev.\  D {\bf 77}, 043515 (2008)
 [arXiv:0709.1485 [astro-ph]].

\bibitem{Sandin:2008db}
 F.~Sandin and P.~Ciarcelluti,
 Astropart.\ Phys.\  {\bf 32}, 278 (2009)
 [arXiv:0809.2942 [astro-ph]].

\bibitem{McCullough:2010ai}
 M.~McCullough and M.~Fairbairn,
 Phys.\ Rev.\  D {\bf 81}, 083520 (2010)
 [arXiv:1001.2737 [hep-ph]].

\bibitem{Kouvaris:2010vv}
 C.~Kouvaris and P.~Tinyakov,
 Phys.\ Rev.\  D {\bf 82}, 063531 (2010)
 [arXiv:1004.0586 [astro-ph.GA]].

\bibitem{Casanellas:2010sj}
 J.~Casanellas and I.~Lopes,
 arXiv:1002.2326 [astro-ph.CO].

\bibitem{deLavallaz:2010wp}
 A.~de Lavallaz and M.~Fairbairn,
 Phys.\ Rev.\  D {\bf 81}, 123521 (2010)
 [arXiv:1004.0629 [astro-ph.GA]].

\bibitem{Kouvaris:2010jy}
 C.~Kouvaris and P.~Tinyakov,
 Phys.\ Rev.\  D {\bf 83}, 083512 (2011)
 [arXiv:1012.2039 [astro-ph.HE]].

\bibitem{Ciarcelluti:2010ji}
 P.~Ciarcelluti and F.~Sandin,
 Phys.\ Lett.\  B {\bf 695}, 19 (2011)
 [arXiv:1005.0857 [astro-ph.HE]].

\bibitem{Kouvaris:2011fi}
 C.~Kouvaris and P.~Tinyakov,
 Phys.\ Rev.\ Lett.\  {\bf 107}, 091301 (2011)
 [arXiv:1104.0382 [astro-ph.CO]].

\bibitem{McDermott:2011jp}
 S.~D.~McDermott, H.~B.~Yu and K.~M.~Zurek,
 arXiv:1103.5472 [hep-ph].

\bibitem{Fornengo:2011sz}
  N.~Fornengo, P.~Panci, M.~Regis,
  arXiv:1108.4661 [hep-ph].

 \bibitem{virial}
 H.~H.~Schmidtke,
 Theoret. chim. Acta {\bf 8}, 376-382 (1967)
 
 
\bibitem{Kargaltsev:2003eb}
 O.~Kargaltsev, G.~G.~Pavlov and R.~W.~Romani,
 Astrophys.\ J.\  {\bf 602}, 327 (2004)
 [arXiv:astro-ph/0310854].

\bibitem{Mignani:2008jr}
 R.~P.~Mignani, G.~G.~Pavlov and O.~Kargaltsev,
 arXiv:0805.2586 [astro-ph].

  
\bibitem{Brayeur:2011yw}
  L.~Brayeur, P.~Tinyakov,
  [arXiv:1111.3205 [astro-ph.CO]].  

\bibitem{Randall:2007ph}
  S.~W.~Randall, M.~Markevitch, D.~Clowe, A.~H.~Gonzalez, M.~Bradac,
  Astrophys.\ J.\  {\bf 679}, 1173-1180 (2008).
  [arXiv:0704.0261 [astro-ph]].  
  
    \end{thebibliography}
      \end{document}